\documentclass[aps,reprint,amsmath,amssymb,onecolumn]{revtex4-1}

\usepackage{graphicx,epsfig,epsf,color,hhline,import}
\usepackage{dcolumn}
\usepackage{bm}
\usepackage{tensor,pbox}
\usepackage{tikz}
\usepackage{epstopdf}
\usepackage{amsthm}
\usepackage{bbold}
\usepackage[ruled,vlined]{algorithm}
\usepackage{changes}
\usepackage{cancel}
\usepackage{color}
\usepackage{url}
\usepackage[hidelinks]{hyperref}
\usepackage{ytableau}
\usepackage{enumerate}
\usepackage{mathdots}
\usepackage{braket}
\usepackage{MnSymbol}
\usepackage{mathtools}

\makeatletter
\renewcommand{\p@subsection}{}
\renewcommand{\p@subsubsection}{}
\makeatother

\newcommand{\one}{{\mathbb{1}}}
\newcommand{\tr}{{\rm tr}}

\newcommand{\trb}[1]{\text{tr}\left[{#1}\right]}

\newcommand{\ii}{\ensuremath{ \mathrm{i\,} }}
\newcommand{\e}[1]{\ensuremath{  \operatorname{e}^{#1}}}

\newcommand{\Ns}{{N_\text{S}}}
\newcommand{\Nc}{{N_\text{C}}}

\newcommand{\vecb}{{\operatorname{vecb}}}
\newcommand{\vecc}{{\operatorname{vec}}}

\definecolor{bostonuniversityred}{rgb}{0.8, 0.0, 0.0}

\begin{abstract}
	The quantum Fisher information matrix is a central object in multiparameter quantum estimation theory. It is usually challenging to obtain analytical expressions for it because most calculation methods rely on the diagonalization of the density matrix. In this paper, we derive general expressions for the quantum Fisher information matrix which bypass matrix diagonalization and do not require the expansion of operators on an orthonormal set of states. Additionally, we can tackle density matrices of arbitrary rank. The methods presented here simplify analytical calculations considerably when, for example, the density matrix is more naturally expressed in terms of non-orthogonal states, such as coherent states. Our derivation relies on two matrix inverses which, in principle, can be evaluated analytically even when the density matrix is not diagonalizable in closed form.
	We demonstrate the power of our approach by deriving novel results in the timely field of discrete quantum imaging: the estimation of positions and intensities of incoherent point sources. We find analytical expressions for the full estimation problem of two point sources with different intensities, and for specific examples with three point sources. We expect that our method will become standard in quantum metrology.
\end{abstract}
\begin{document}
	\title{General expressions for the quantum Fisher information matrix \\ with applications to discrete quantum imaging}
	\author{Lukas J.~Fiderer$^{1,2}$, Tommaso Tufarelli$^1$, Samanta Piano$^3$, Gerardo Adesso$^1$}
	\affiliation{
		$^1$School of Mathematical Sciences,  University of Nottingham, University Park, Nottingham NG7 2RD, United Kingdom\\
		$^2$Institute for Theoretical Physics, University of Innsbruck, 6020 Innsbruck, Austria\\
		$^3$ Manufacturing Metrology Team, Faculty of Engineering, University of Nottingham, Nottingham NG7 2RD, United Kingdom
	}
	\maketitle
	
	\onecolumngrid
	
	Quantum metrology deals with the estimation of unknown parameters from the measurement outcomes of quantum experiments. Naturally, the goal is to design experiments and data-analysis strategies such that the uncertainty in the estimated parameters is minimized. One of the appeals of quantum metrology is that it promises reduced uncertainties compared to what is possible with comparable classical resources \cite{giovannetti_quantum_2004,paris_quantum_2009,giovannetti_advances_2011, pezze2018quantum, braun2018quantum}. In order to understand and quantify the advantage offered by quantum metrology, we typically calculate the quantum Fisher information (QFI) or, in the  general case of simultaneous estimation of multiple parameters, the QFI matrix (QFIM). The inverse of the QFIM yields the quantum Cram\'er--Rao bound (QCRB), a lower bound on the uncertainty of any unbiased estimator of the parameters. In particular, analytical solutions for the QFIM are desirable as they provide valuable insights into how the estimation error depends on and scales with tunable parameters. Since the QFIM is a function of the density matrix, existing general approaches to compute the QFIM conventionally assume that the density matrix is expressed with respect to an orthogonal basis \cite{safranek2018simple}, or even start with a density matrix in its diagonalized form \cite{liu2019quantum}. However, analytical matrix diagonalization is typically limited to low dimensions. On top of that, the density matrix often has a natural and elegant representation with respect to some non-orthogonal basis. In such cases it is not only cumbersome to choose a suitable orthogonal basis and to expand the density matrix in such a basis, but it also complicates the further computation of the QFIM. 
	
	In this work, we address these problems and provide a general and efficient method to analytically compute the QFIM. Our approach is based on a new formal solution for the QFIM. Compared to previous methods, our solution relies on a non-orthogonal basis approach \cite{napoli2019towards, genoni2019non} which allows us to express all matrices with respect to an arbitrary, possibly non-orthogonal basis. Further, our solution does not rely on matrix diagonalization but on matrix inversion.  Our results improve over the analysis of Refs.~\cite{napoli2019towards, genoni2019non} by providing an expression for the QFIM which relies on the general solution of the associated Lyapunov equations and, thus, avoids solving the Lyapunov equations for each parameter separately. In comparison with Ref.~\cite{bisketzi2019quantum}, our expressions for the QFIM are general and do not depend on particular properties of the problem under consideration.
	
	There are many situations in quantum metrology where the natural representation of $\rho$ involves a non-orthogonal set of states, e.g., when coherent states are involved (such as for noisy Schr\"odinger cat states or entangled coherent states \cite{genoni2019non}). In this work, we demonstrate the power of our approach by deriving novel analytical results in the field of discrete quantum imaging, a  rapidly developing branch of quantum metrology \cite{tsang2019resolving}. For the problem of imaging two incoherent point sources of monochromatic light in the paraxial regime, we provide a complete solution for the QFIM when all parameters, i.e., the spatial locations and relative intensity of the sources, are estimated simultaneously. Our method further allows us to derive analytical results for some special arrangements of three point sources, providing new insights into the imaging of multiple sources.
	
	The paper is organized as follows. In Section \ref{sec:prelim}, we provide a precise formulation of the problem and discuss existing approaches to compute the QFIM. In Section~\ref{sec:QFIM_for_general_bases}, we present our formal solution for multiparameter quantum estimation with respect to non-orthogonal bases.  In Section~\ref{sec:discrete_q_imaging}, we introduce quantum imaging and present our results for the QFIM for two and three point sources.
	In Section~\ref{sec:discussion}, we discuss the computational requirements of our method and close the paper with some concluding remarks.
	
	\section{Preliminaries} \label{sec:prelim}
	
	Let us consider the general case of simultaneously estimating $n$ parameters, $ \boldsymbol{\theta} =(\theta_{1}, \dotsc, \theta_{n})$. Further, let $ \hat{\boldsymbol{\theta}}=(\hat\theta_{1}, \dotsc, \hat\theta_{n})$ be an estimator of $ \boldsymbol{\theta}$, with $ \hat\theta_{\mu} $ the estimator of $ \theta_{\mu} $. The uncertainty in the estimator $ \hat{\boldsymbol{\theta}} $ can be characterized by the covariance matrix $\text{Cov}\left(\hat{\boldsymbol{\theta}}\right)$ and is lower bounded by the QCRB \cite{helstrom_quantum_1976,braunstein_statistical_1994}
	\begin{align}
	\text{Cov}\left(\hat{\boldsymbol{\theta}}\right)\geq \frac{1}{M}\boldsymbol{H}^{-1}, \label{eq:QCRB}
	\end{align}
	with $M$ the number of repetitions of the experiment (i.e. the statistical ensemble size), and $\boldsymbol{H}$ the quantum Fisher information matrix (QFIM). We use the notation that vectors and matrices are printed in bold while scalars and operators are not. Further, $\theta_j$ denotes the $j$th coefficient of vector $\boldsymbol{\theta}$ and $H_{\mu,\nu}$ the $(\mu,\nu)$th coefficient of matrix $\boldsymbol{H}$. The QCRB represents a matrix inequality in the sense that $\text{Cov}\left(\hat{\boldsymbol{\theta}}\right)- \frac{1}{M}\boldsymbol{H}^{-1}$ has to be a positive semi-definite matrix. Instead of the matrix inequality \eqref{eq:QCRB}, one often considers a lower bound on the summed variances of each $ \hat{\theta}_{j}$ which is obtained by taking the trace on both sides of Eq.~\eqref{eq:QCRB}.
	
	For a parameter-dependent density operator $\rho$, the coefficients of the QFIM $\boldsymbol{H}$ are defined as
	\begin{align}
	H_{\mu,\nu}=\tr\left(L_\mu\partial_\nu\rho\right),\label{eq:qfi}
	\end{align}
	where $\partial_{\nu}= \partial/\partial{\theta_{\nu}}$ is shorthand for the derivative by $\theta_{\nu}$, and $ L_{\mu}$  is the symmetric logarithmic derivative (SLD) for the parameter $\theta_{\mu} $. In general, $\rho$, the SLDs, and the QFIM depend on the parameters $\boldsymbol{\theta}$, however, we drop any parameter dependence in our notation for the sake of brevity. The SLDs are defined via Lyapunov equations
	\begin{align}
	2\partial_{\mu}\rho= \rho L_{\mu}+L_{\mu}\rho, \label{eq:Lyapunov}
	\end{align}
	and by inserting Eq.~\eqref{eq:Lyapunov} for the derivative in Eq.~\eqref{eq:qfi} it becomes apparent that $\boldsymbol{H}$ is a symmetric matrix. 
	
	Analytical solutions for the QFIM are usually obtained based on the following classical result: for any density operator $ \rho $ with nonzero eigenvalues, a formal solution for Eq.~\eqref{eq:Lyapunov} is given by
	\cite{bellman1997introduction}
	\begin{align}
	L_{\mu}=2\int_{0}^{\infty} \e{-\rho s}\left(\partial_{\mu}\rho\right) \e{-\rho s} \text{d}s.\label{eq:sld_formal_solution}
	\end{align}
	Similarly, the QFIM can be written as \cite{paris_quantum_2009}
	\begin{align}
	H_{\mu,\nu}= 2\int_{0}^{\infty} \tr\left[ \e{-\rho s}\left(\partial_{\mu}\rho\right) \e{-\rho s} \left(\partial_{\nu}\rho\right)\right] \text{d}s.\label{eq:qfi_formal_integral_solution}
	\end{align}
	The most common method to solve the integral in Eq.~\eqref{eq:sld_formal_solution} relies on expanding $ \rho $ in its eigendecomposition $ \rho=\sum_{j=1}^d\lambda_{j}\ket{\lambda_{j}}\bra{\lambda_{j}} $, with nonzero eigenvalues $\lambda_j$ and eigenvectors $ \ket{\lambda_j} $.  Inserting the solution of the integral \eqref{eq:sld_formal_solution} into Eq.~\eqref{eq:qfi} yields the QFIM	
	\begin{align}
	H_{\mu,\nu}= 2\sum_{j, k=1}^{d} \frac{\operatorname{Re}\left(\braket{\lambda_{j}|\partial_{\mu}\rho|\lambda_{k}}\braket{\lambda_{k}|\partial_{\nu}\rho|\lambda_{j}}\right)}{\lambda_{j}+\lambda_{k}}; \label{eq:qfi_diagonalization}
	\end{align}
	see Ref.~\cite{liu2019quantum} for an overview of analytical results for the QFIM based on the eigendecomposition of $ \rho $, including formulas which hold if $\rho$ does not have full rank. In particular, Eq.~\eqref{eq:qfi_diagonalization} holds also for non-full-rank $\rho$ if one restricts the summation to indices for which $\lambda_l+\lambda_m>0$ holds.

	However, the density operator $ \rho $ is usually not given in its eigendecomposition, and the diagonalization of $ \rho $ is known to be a hard problem; it requires solving the characteristic equation, i.e., a polynomial equation of order $d$. Abel's impossibility theorem \cite{abel1826beweis} states that algebraic solutions for polynomial equations with arbitrary coefficients are impossible for dimension $ d\geq 5 $. Therefore, finding analytical solutions for the QFIM by expanding $ \rho $ in its eigenbasis usually works only for low dimensions or some special cases.
	
	An alternative to the formal solution \eqref{eq:sld_formal_solution} is to expand the Lyapunov equations \eqref{eq:Lyapunov} in an orthonormal basis and to vectorize the corresponding matrix equation, which is a well-known approach in the mathematical literature \cite{laub2005matrix} but has received little attention in quantum metrology \cite{safranek2018simple, bisketzi2019quantum}. Using vectorization, the Lyapunov equations are transformed into a linear system which can be solved without diagonalizing $ \rho $. In particular, solving a $d$-dimensional linear system can in principle be done analytically for any finite $d$ and does not suffer from the same limitations as solving polynomial equations of order $d$, i.e., computing eigenvalues of $\rho$. If there exists an orthonormal basis, i.e, an orthonormal set of states supporting $\rho$ and its derivatives (by the parameters of interest) such that the density matrix $ \boldsymbol{\rho} $ (formed by the coefficients of $\rho$ in that basis) has full rank, \v{S}afr\'anek gives the following formal solution for the QFIM \cite{safranek2018simple}
	\begin{align}
	H_{\mu,\nu}= 2\, \vecc\left(\boldsymbol{\partial_{\mu}\rho}\right)^\dagger\left(\overline{\boldsymbol{\rho}}\otimes \one + \one \otimes \boldsymbol{\rho}\right)^{-1}\vecc \left(\boldsymbol{\partial_{\nu}\rho}\right), \label{eq:qfi_inverse}
	\end{align}
	where $\boldsymbol{\partial_{\mu}\rho}$ ($\boldsymbol{\partial_{\nu}\rho}$) is the matrix formed by the coefficients of $\partial_{\mu}\rho$ ($\partial_{\nu}\rho$) expanded in the orthonormal basis, $\one$ denotes the identity matrix of the same dimension as $\boldsymbol{\rho}$, $\overline{\boldsymbol{A}}$ denotes the complex conjugate of $\boldsymbol{A}$, $\boldsymbol{A}^\dagger$ denotes the conjugate transpose of $\boldsymbol{A}$, and $\operatorname{vec}(\boldsymbol{A})$ is defined as a column vector obtained by concatenating the columns $\boldsymbol{a}_j$ of $\boldsymbol{A}$,
	\begin{align}
	\operatorname{vec}(\boldsymbol{A})=\begin{pmatrix}
	\boldsymbol{a}_1 \\
	\boldsymbol{a}_2 \\
	\vdots \\
	\boldsymbol{a}_d
	\end{pmatrix}. \label{eq:vec_def}
	\end{align}
	In the derivation of \v{S}afr\'anek's formula, Eq.~\eqref{eq:qfi_inverse}, the inverse matrix in Eq.~\eqref{eq:qfi_inverse} is part of the formal solution of the Lyapunov equations. Compared to Eq.~\eqref{eq:qfi_diagonalization}, \v{S}afr\'anek's formula has the advantage that it does not rely on matrix diagonalization but instead uses the inverse of a $d^2\times d^2$ matrix; matrix inversion is equivalent to solving a linear system and thus does not share the limitations of analytical matrix diagonalization.
	
	A drawback of Eq.~\eqref{eq:qfi_inverse} is that it requires $\boldsymbol{\rho}$ to be invertible, i.e., $\rho$ needs to be given with respect to an orthogonal basis such that its matrix has full rank. If $\boldsymbol{\rho}$ does not have full rank and, thus, is not invertible, one can 
	replace $ \boldsymbol{\rho} $ with an invertible matrix $\boldsymbol{\rho}_s = (1-s)\boldsymbol{\rho}+s/d\,\one$ such that the QFIM is then given by \cite{safranek2018simple}
	\begin{align}
	H_{\mu,\nu}= 2\lim\limits_{s\rightarrow 0}\, \vecc\left(\boldsymbol{\partial_{\mu}\rho}_s\right)^\dagger\left(\overline{\boldsymbol{\rho}_s}\otimes \one + \one \otimes \boldsymbol{\rho}_s\right)^{-1}\vecc \left(\boldsymbol{\partial_{\nu}\rho}_s\right). \label{eq:qfi_inverse_s}
	\end{align}
	However, Eq.~\eqref{eq:qfi_inverse_s} involves additional analytical computations, and if the dimension $d$ of $\boldsymbol{\rho}$ is much larger than its rank, the matrix to be inverted is much larger than necessary. Additionally, $\rho$ is often given with respect to non-orthogonal states which form a basis spanning the support of $\rho$. Then, the matrix of $\rho$ with respect to this non-orthogonal basis has a compact form and full rank. In such cases, it is nontrivial to find a suitable orthogonal basis which spans only the support of $\rho$. Instead, one often has to rely on bases for the whole Hilbert space which leads to an inefficient representation of $\rho$ if the rank of $\rho$ is smaller than the dimension of the Hilbert space.
	
	In the following, we address these problems by deriving a general formal solution for the QFIM using a non-orthogonal basis approach \cite{napoli2019towards, genoni2019non}. Similarly to Eq.~\eqref{eq:qfi_inverse}, our solution does not rely on matrix diagonalization but on matrix inversion, and we will show that it can be seen as a generalization of \v{S}afr\'anek's formula to non-orthogonal bases.
	
	\section{The Quantum Fisher information matrix for general bases} \label{sec:QFIM_for_general_bases}
	In this section we will present our general expressions for the QFIM and discuss some special cases. The derivation (see Appendix \ref{app:derivation}) uses block-vectorization, a variation of standard vectorization, which allows us to separate a basis into different parts, corresponding to different subspaces. In this way, we define all relevant matrices on their support such that they are invertible. 
	
	\subsection{QFIM for general bases} \label{sec:general_result}
	
	Our general solution for the QFIM relies only on one assumption, which is that the density operator $\rho$ is given with respect to a $ d $-dimensional basis $\mathcal{B}=\left\{\ket{\psi_j}\right\}_{j=1}^{d}$, where $\mathcal{B}$ is a set of linearly independent states $\ket{\psi_j}$ spanning the support of $\rho$. Note that we do not assume that the basis $ \mathcal{B} $ is orthogonal. Then, $ \rho=\sum_{j,k=1}\rho^{\mathcal{B}}_{j,k}\ket{\psi_j}\bra{\psi_k}$ can be represented by a full-rank hermitian matrix $\boldsymbol{\rho}^{\mathcal{B}}$ with coefficients $\rho^{\mathcal{B}}_{j,k}$. In the following, we write $\boldsymbol{A}^{\mathcal{B}}$ for the matrix representation of $A$ with respect to the basis $\mathcal{B}$.
	
	Matrix equations can be rewritten for matrices which are given with respect to a general, non-orthogonal basis $ \mathcal{B} $ using the Gramian $ G^{\mathcal{B}}_{j,k}=\braket{\psi_{j}|\psi_{k}} $, defined with respect to the basis $ \mathcal{B}$. Then, for matrices $ \boldsymbol{A} $ and $ \boldsymbol{B} $, defined with respect to an orthogonal basis, we have the following replacements for the matrix product and the trace operation \cite{genoni2019non}
	\begin{align}
	\boldsymbol{AB} &\rightarrow \boldsymbol{A}^{\mathcal{B}}\boldsymbol{G}^{\mathcal{B}} \boldsymbol{B}^{\mathcal{B}}, \label{eq:replace1}\\
	\trb{\boldsymbol{A}} &\rightarrow \trb{\boldsymbol{A}^{\mathcal{B}}\boldsymbol{G}^{\mathcal{B}}}. \label{eq:replace2}
	\end{align}
	Clearly, if $\mathcal{B}$ is orthonormal, then $ \boldsymbol{G}^{\mathcal{B}}=\one $ and we retrieve the standard matrix operations.
	With this, the Lyapunov equation for the parameter $ \theta_{\mu} $ becomes
	\begin{align}
	2 \left(\boldsymbol{\partial_\mu\rho}\right)^{\mathcal{B}_\mu} = \boldsymbol{L}^{\mathcal{B}_\mu}_\mu \boldsymbol{G}^{\mathcal{B}_\mu} \boldsymbol{\rho}^{\mathcal{B}_\mu}+\boldsymbol{\rho}^{\mathcal{B}_\mu} \boldsymbol{G}^{\mathcal{B}_\mu} \boldsymbol{L}^{\mathcal{B}_\mu}_\mu, \label{eq:sld_eqs}
	\end{align}
	where $\mathcal{B}_\mu$ denotes a basis which spans the support of $\rho$ and $\partial_\mu\rho$, and $\boldsymbol{G}^{\mathcal{B}_\mu}$ is the Gramian of $\mathcal{B}_\mu$.

	Let us denote the $(i,j)$th matrix block of a matrix $\boldsymbol{A}$ as $\boldsymbol{A}_{ij}$. In comparison, $A_{i,j}$ denotes a matrix coefficient. In the following, all matrices are divided in 4 blocks as follows: the $\boldsymbol{A}_{11}$ block is always of size $\left|\mathcal{B}\right|\times \left|\mathcal{B}\right|$ where $|\bullet|$ denotes cardinality and $ \mathcal{B} $ the given basis spanning the support of $ \rho $. The remaining blocks complete the total matrix. For example, we have
	\begin{align}
	\boldsymbol{\rho}^{\mathcal{B}_{\mu}}=\begin{bmatrix}
	\boldsymbol{\rho}^{\mathcal{B}} & 0 \\
	0 & 0
	\end{bmatrix}, \label{eq:rhoblock}
	\end{align}
	where the zeros denote blocks of zeros such that $\boldsymbol{\rho}^{\mathcal{B}_{\mu}}$ is of size $\left|\mathcal{B}_{\mu}\right|\times \left|\mathcal{B}_{\mu}\right|$.
	If the derivative of $\partial_\mu\rho$ lies within the support of $\rho$, all but the $\boldsymbol{\rho}^{\mathcal{B}_\mu}_{11}=\boldsymbol{\rho}^{\mathcal{B}}$ block vanish.

	In Appendix \ref{app:derivation}, we use block-vectorization to derive the following general solution for the SLD
	\begin{align}
	\boldsymbol{L}_{\mu}^{\mathcal{B}_\mu} =2\begin{pmatrix}
	\operatorname{mat}\left(\boldsymbol{D}^{-1}\vecc\left[\left(\boldsymbol{\partial_\mu\rho}\right)^{\mathcal{B}_\mu} _{11}-\boldsymbol{E}-\boldsymbol{E}^\dagger\right]\right) & \boldsymbol{C}^{-1}\left(\boldsymbol{\partial_\mu\rho}\right)^{\mathcal{B}_\mu}_{12}\\
	\left(\boldsymbol{\partial_\mu\rho}\right)^{\mathcal{B}_\mu}_{21} \left(\boldsymbol{C}^{-1}\right)^\dagger & 0\\
	\end{pmatrix}, \label{eq:SLDformal}
	\end{align}
	where we defined the matrices
	\begin{align}
	\boldsymbol{C}&=\boldsymbol{\rho}_{11}^{\mathcal{B}_\mu}\boldsymbol{G}_{11}^{\mathcal{B}_\mu}, \label{eq:c}\\
	\boldsymbol{D}&=\one_{11}^{\mathcal{B}_\mu}\otimes  \boldsymbol{C} + \overline{ \boldsymbol{C}}\otimes \one_{11}^{\mathcal{B}_\mu}, \label{eq:d}\\
	\boldsymbol{E}&=\boldsymbol{C}^{-1}\left(\boldsymbol{\partial_\mu\rho}\right)^{\mathcal{B}_\mu}_{12} \boldsymbol{G}_{21}^{\mathcal{B}_\mu}\boldsymbol{\rho}_{11}^{\mathcal{B}_\mu},\label{eq:e}
	\end{align}
	and where $\operatorname{mat}(\cdot)$ is defined to take a vector of length $n^2$ and rearrange its coefficients to a matrix of size $n\times n$ by inserting the first $n$ coefficients of the vector to the first column of the matrix, the next $n$ coefficients to the second column, and so forth; for example for $n=2$
	\begin{align}
	\boldsymbol{a}=\begin{pmatrix}
	a_1\\a_2\\a_3\\a_4
	\end{pmatrix},\qquad \operatorname{mat}\left(\boldsymbol{a}\right)=\begin{pmatrix}
	a_1 & a_3 \\
	a_2 & a_4
	\end{pmatrix}. \label{eq:mat_example}
	\end{align}
	
	All that remains to be done is to use Eq.~\eqref{eq:SLDformal} to calculate the QFIM which, using our replacement rules (\ref{eq:replace1}-\ref{eq:replace2}), is given by
	\begin{align}
	H_{\mu,\nu}=\tr\left(\boldsymbol{L}^{\mathcal{B}_{\mu,\nu}}_\mu  \boldsymbol{G}^{\mathcal{B}_{\mu,\nu}} \left(\boldsymbol{\partial_\nu\rho}\right)^{\mathcal{B}_{\mu,\nu}}  \boldsymbol{G}^{\mathcal{B}_{\mu,\nu}}\right), \label{eq:qfim_nonortho}
	\end{align}
	where $\mathcal{B}_{\mu,\nu}$ is a basis spanning the support of $\rho$, $\partial_\mu\rho$, and $\partial_\nu\rho$. The matrix $\boldsymbol{L}^{\mathcal{B}_{\mu,\nu}}_\mu$ is obtained by padding the matrix $\boldsymbol{L}^{\mathcal{B}_{\mu}}_\mu$ in Eq.~\eqref{eq:SLDformal} with zeros, i.e., $(\boldsymbol{L}^{\mathcal{B}_{\mu,\nu}}_\mu)_{i,j}=0$ if one or both of the states $\ket{\psi_i}$ and $\ket{\psi_j}$ lie outside $\mathcal{B}_{\mu}$.

	Eqs.~\eqref{eq:SLDformal} and \eqref{eq:qfim_nonortho} represent our general solution for the QFIM. The only nontrivial calculations are to find the inverses of $\boldsymbol{C}$ and $\boldsymbol{D}$, as given by Eqs.~\eqref{eq:c} and \eqref{eq:d}. Similar methods have been used in Ref.~\cite{bisketzi2019quantum}, although our result is more general because the result of Ref.~\cite{bisketzi2019quantum} depends on particular properties of the problem considered in that work.
	
	We proceed by expressing the compatibility conditions \cite{ragy2016compatibility} with respect to a non-orthogonal basis. This facilitates the evaluation of the compatibility conditions if the QFIM is computed using Eqs.~\eqref{eq:SLDformal} and \eqref{eq:qfim_nonortho}, and in Section \ref{sec:discrete_q_imaging} we will discuss the compatibility conditions in the context of discrete quantum imaging.
	
	\subsection{Compatibility conditions} \label{sec_compatibility}
	When considering a ``multiparameter scenario'', i.e., the simultaneous estimation of $n>1$ parameters, an interesting question is how it compares to an (overly optimistic) ``separate scenario'' where one assumes that each parameter can be estimated independently, i.e., that there is an estimation scheme for each parameter, and for each individual estimation scheme one assumes that (i) all other parameters are known and (ii) the same resources are available as for the multiparameter scenario. While such separate scenario is an idealization which usually cannot be implemented (see Refs.~\cite{tsang2020quantum, suzuki2020quantum} for a rigorous treatment of quantum parameter estimation with nuisance parameters), it represents a useful benchmark and performs at least as well as the multiparameter scenario. If the so-called {\em compatibility conditions} are fulfilled, the multiparameter scenario matches the performance of the separate scenario (while using only the resources of one of the estimation schemes) \cite{ragy2016compatibility}. The compatibility conditions consist of (i) the \textit{commutation condition}, $\Gamma_{\mu,\nu}=0$ for all $\mu\neq\nu$, where
	\begin{align} \label{eq:gamma}
	\Gamma_{\mu,\nu}
	&=\operatorname{Im}\left(\tr\left[\boldsymbol{\rho}^{\mathcal{B}_{\mu,\nu}} \boldsymbol{G}^{\mathcal{B}_{\mu,\nu}} \boldsymbol{L}^{\mathcal{B}_{\mu,\nu}}_\mu \boldsymbol{G}^{\mathcal{B}_{\mu,\nu}} \boldsymbol{L}^{\mathcal{B}_{\mu,\nu}}_\nu\boldsymbol{G}^{\mathcal{B}_{\mu,\nu}}\right]\right),
	\end{align}
	(ii) the \textit{independence condition}, $H_{\mu,\nu}=0$ for all $\mu\neq\nu$, and (iii) the \textit{initial-state condition}, i.e., that there exists a single initial probe state which is optimal for every estimation scheme in the separate scenario.
	
	Note that the QFIM $\boldsymbol{H}$ and $\boldsymbol{\Gamma}$ can be seen as real and imaginary part of the same quantity since 
	\begin{align}
	H_{\mu,\nu}=\operatorname{Re}\left(\tr\left[\boldsymbol{\rho}^{\mathcal{B}_{\mu,\nu}} \boldsymbol{G}^{\mathcal{B}_{\mu,\nu}} \boldsymbol{L}^{\mathcal{B}_{\mu,\nu}}_\mu \boldsymbol{G}^{\mathcal{B}_{\mu,\nu}} \boldsymbol{L}^{\mathcal{B}_{\mu,\nu}}_\nu\boldsymbol{G}^{\mathcal{B}_{\mu,\nu}}\right]\right).
	\end{align}

	If the commutation condition is fulfilled, there exist optimal measurements \footnote{In general, joint measurements of multiple copies of $\rho(\boldsymbol{p},\boldsymbol{r})$ might be necessary to asymptotically saturate the QCRB \cite{ragy2016compatibility}. } such that the QCRB can be saturated (although the estimators of the parameters might not be independent). It is possible to find independent estimators for the parameters if the independence condition is fulfilled, i.e., if the off-diagonal coefficients of the QFIM vanish. If, in addition to the commutation and independence conditions, the initial-state condition is fulfilled, an optimal multiparameter scenario can be constructed which matches the performance of the corresponding separate scenario.
	
	The remainder of this section is devoted to showing how to recover special cases of relevance from our general results.
	
	\subsection{Retrieving \v{S}afr\'anek's formula as a special case}
	In order to retrieve \v{S}afr\'anek's formula, as given in Eq.~\eqref{eq:qfi_inverse}, from our general solution, we have to make the additional assumption that the basis $\mathcal{B}$, which spans the support of $\rho$, is orthogonal, and that the derivatives $\partial_\mu\rho$ are supported by $\mathcal{B}$. Then, all matrices can be expressed with respect to $\mathcal{B}$, and $\boldsymbol{G}^{\mathcal{B}}=\one$. This means that $\boldsymbol{C}=\boldsymbol{\rho}^\mathcal{B}$, $\boldsymbol{D}=\one^{\mathcal{B}}\otimes \boldsymbol{\rho}^\mathcal{B}+ \overline{\boldsymbol{\rho}^\mathcal{B}}\otimes \one^{\mathcal{B}}$, and $\boldsymbol{E}$ vanishes because $\mathcal{B}_\mu=\mathcal{B}$ for all $\mu$. The SLD is then given by $\boldsymbol{L}_{\mu}^{\mathcal{B}} =2\operatorname{mat}\left(\boldsymbol{D}^{-1}\vecc\left[\boldsymbol{\partial_\mu\rho}^{\mathcal{B}}\right]\right) $ and the QFIM is found to be
	\begin{align}
	H_{\mu,\nu}&=\tr\left(\boldsymbol{L}^{\mathcal{B}}_\mu  \boldsymbol{\partial_\nu\rho}^{\mathcal{B}}\right) \\
	&=2\,\tr\left(\operatorname{mat}\left[\boldsymbol{D}^{-1}\vecc\left(\boldsymbol{\partial_\mu\rho}^{\mathcal{B}}\right) \right] \boldsymbol{\partial_\nu\rho}^{\mathcal{B}} \right) \label{eq:line1}\\
	&=2\,\vecc\left(\operatorname{mat}\left[\boldsymbol{D}^{-1}\vecc\left(\boldsymbol{\partial_\mu\rho}^{\mathcal{B}}\right)\right]  \right)^\dagger \vecc\left(\boldsymbol{\partial_\nu\rho}^{\mathcal{B}}\right) \label{eq:line2} \\
	&=2\,\vecc\left(\boldsymbol{\partial_\mu\rho}^{\mathcal{B}}\right)^\dagger \left(\one^{\mathcal{B}}\otimes \boldsymbol{\rho}^\mathcal{B}+ \overline{\boldsymbol{\rho}^\mathcal{B}}\otimes \one^{\mathcal{B}}\right)^{-1} \vecc\left(\boldsymbol{\partial_\nu\rho}^{\mathcal{B}}\right),\label{eq:line3}
	\end{align}
	where we used $\trb{\boldsymbol{AB}}=\vecc\left(\boldsymbol{A}\right)^\dagger\vecc\left(\boldsymbol{B}\right)$ to get to Eq.~\eqref{eq:line2}, and we used that the inverse of an invertible hermitian matrix is hermitian to find Eq.~\eqref{eq:line3} which is indeed identical with Eq.~\eqref{eq:qfi_inverse}. Also, since Eq.~\eqref{eq:qfi_diagonalization} can be seen as a special case of Eq.~\eqref{eq:qfi_inverse} \cite{safranek2018simple}, it is a special case of our general solution as well.
	\subsection{Unitary parameterization with commuting generators}
	Let us introduce a simple unitary parameterization model using operator equations. The parameter-dependent density operator is given by $\rho=U(\boldsymbol{\theta})\rho_0 U^\dagger(\boldsymbol{\theta})$, with $\rho_0$ an parameter-independent initial state, $U(\boldsymbol{\theta})=\exp\left(-\ii\sum_{\mu=1}^{n}K_\mu \theta_\mu\right)$ a unitary operator which encodes the parameter dependence into $\rho_0$, and we assume that the generators $K_\mu$ commute.
	
	It is then easy to show that the QFIM can be written as 
	\begin{align}
	H_{\mu,\nu} &= \tr\left(L'_\mu\rho'_\nu\right), \label{eq:qfim_unitary}
	\end{align}
	where $\rho'_\mu=\left[K_\mu,\rho_0\right]$ replaces the derivative when compared with Eq.~\eqref{eq:qfi}, and $L'_\mu$ is given via
	\begin{align}
	2 \rho'_\mu &= \rho_0 L'_\mu+L'_\mu\rho_0. \label{eq:Lyapunov_unitary}
	\end{align} 
	We can solve Eq.~\eqref{eq:Lyapunov_unitary} with respect to a general basis in the same way as we solved the Lyapunov equation above. In particular, Eq.~\eqref{eq:SLDformal} becomes a solution for ${\boldsymbol{L'}_\mu}^{\mathcal{B}_\mu}$ if we replace $\left(\boldsymbol{\partial_\mu\rho}\right)^{\mathcal{B}_\mu}$ with ${\boldsymbol{\rho}'_\mu}^{\mathcal{B}_\mu}$ and $\boldsymbol{\rho}^{\mathcal{B}_\mu}$ with $\boldsymbol{\rho}_0^{\mathcal{B}_\mu}$ in all expressions on the right-hand side of Eq.~\eqref{eq:SLDformal} which, together with Eq.~\eqref{eq:qfim_unitary} in matrix form, constitute a solution for the QFIM for the special case of unitary parameterization. Note that this solution does not depend on the parameters $\boldsymbol{\theta}$ [since it does not depend on $U^\dagger(\boldsymbol{\theta})$] but only on the generators $K_\mu$ and the initial state $\rho$.  This is in accordance with known special cases for orthogonal bases \cite{safranek2018simple} and solutions based on matrix diagonalization \cite{liu2019quantum}.
	\subsection{Single-parameter estimation}
	It is worth noticing that the presented method of analytically computing the QFIM also applies to the evaluation of the QFI for single-parameter estimation. The QFI $H$ for estimation of a parameter $\theta_\mu$ is the scalar special case of the QFIM,
	\begin{align}
	H=\tr\left(\boldsymbol{L}^{\mathcal{B}_{\mu}}_\mu  \boldsymbol{G}^{\mathcal{B}_{\mu}} \left(\boldsymbol{\partial_\mu\rho}\right)^{\mathcal{B}_{\mu}}  \boldsymbol{G}^{\mathcal{B}_{\mu}}\right). \label{eq:qfi_nonortho}
	\end{align}
	In particular, avoiding matrix diagonalization and making use of a non-orthogonal basis can be advantageous for computing the QFI as well as the QFIM. In some cases of single-parameter estimation (or even for a few parameters), the analytical computation can be simpler if one solves the linear system [see Eq.~\eqref{eq:app:big_one} in Appendix \ref{app:derivation}] obtained from the Lyapunov equations explicitly for the particular parameter of interest instead of using the general solution \eqref{eq:SLDformal} which holds for any parameter. On the other hand, if one wants to estimate multiple parameters, it will often be more convenient to compute the general solution, i.e., to find the inverses of $\boldsymbol{C}$ and $\boldsymbol{D}$ [Eqs.~(\ref{eq:c}-\ref{eq:d})] instead of solving the linear system \eqref{eq:app:big_one} for each parameter.

	\section{Discrete quantum imaging} \label{sec:discrete_q_imaging}
	In the previous section we derived a formal solution for the QFIM without diagonalizing the density operator $\rho$ and without expressing $\rho$ with respect to an orthogonal set of states. We continue by showing through the technologically relevant example of discrete quantum imaging that our formal solution can be very useful for finding novel analytical expressions for the QFIM.
	
	\subsection{Introduction to quantum imaging}
	Quantum imaging is concerned with using quantum-enhanced detection schemes to image point sources and objects with the highest possible resolution and finding ultimate bounds on the achievable resolution. Potential applications of quantum imaging lie in astronomy, biology, medicine, materials science, and in the semiconductor industry \cite{tsang2019resolving}. 
	
	In 1879, Lord Rayleigh formulated a criterion for the limitations of traditional direct imaging based on classical wave optics: two incoherent point sources cannot be resolved if their separation is significantly smaller than their emission wavelength \cite{rayleigh1879investigations}. Since then, several {\em superresolution} techniques such as fluorescent microscopy \cite{moeckl2014super} have been introduced to overcome Rayleigh's criterion. 
	
	In order to understand the ultimate fundamental limits of imaging, Tsang \textit{et al.}~developed a quantum metrological framework based on the QCRB which relies on a full quantum description of the imaging process \cite{tsang2016quantum, tsang2020quantum}. A key result of Tsang \textit{et al.}~has been that there exist detection schemes such as spatial-mode demultiplexing \cite{tsang2016quantum} which resolve two incoherent point sources with an error independent of their separation and, thus, completely bypass Rayleigh's principle. This has been corroborated by several  proof-of-concept experiments \cite{tang2016fault, yang2016far, paur2016achieving, tham2017beating, parniak2018beating,zhou2019quantum, paur2019reading, boucher2020spatial, zhang2020light, salit2020diffraction}. More detailed studies have shown that for any type of non-adaptive measurement, one- and two-dimensional images of multiple sources in the subdiffraction limit remain unaffected by Rayleigh's criterion only up to the second moment, while for the estimation of higher-order moments a quantum version of Rayleigh's principle reappears \cite{zhou2019modern, peng2020generalization}. Other research has addressed the problem of optimal detection schemes and measurements \cite{yang2019optimal, BonsmaFisher2019realistic} and the implications of practical imperfections such as a misalignment of the detection apparatus \cite{grace2020approaching, dealmeida2020discrimination} or noisy detectors \cite{lupo2020subwavelength}; for a review of recent progress see Ref.~\cite{tsang2019resolving}.
	
	Here, we address the problem of deriving analytical expressions for the QFIM for discrete quantum imaging, i.e., for imaging a discrete set of incoherent point sources.
	In a series of works, analytical expressions for the QFIM have been found for localizing the arbitrary three-dimensional positions of two incoherent point sources of known and possibly unequal brightness \cite{rehacek2017multiparameter,yu2018quantum, napoli2019towards, prasad2019quantum, prasad2020quantum}. Here we go beyond those results by considering the arguably most general detection problem of two incoherent point sources: the joint estimation of their positions and relative intensity, a total of 7 parameters. To the best of our knowledge, we are the first to deliver a fully analytical solution to this problem. Moreover we note that, when considering more than two sources, results have been of numerical nature so far \cite{bisketzi2019quantum}. Using our previously derived expressions for the QFIM, we are able to obtain analytical solutions for some classes of estimation problems involving three incoherent point sources.
	\begin{figure}
		\centering
		\includegraphics[width=0.9\textwidth]{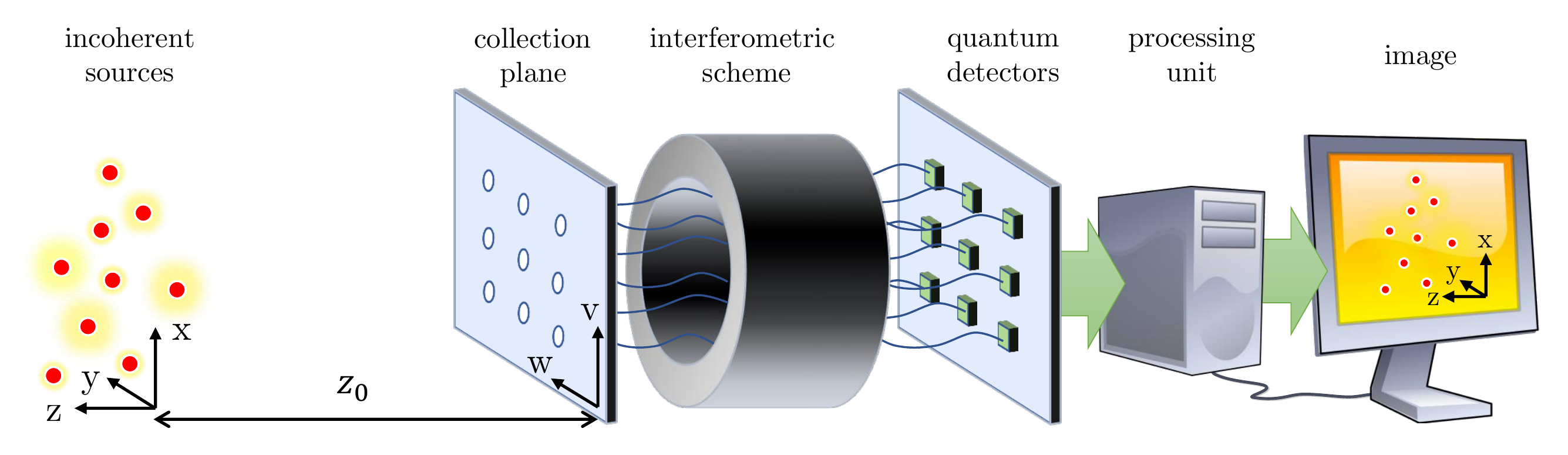}
		\caption{Schematic illustration of discrete quantum imaging. The emitted photons from $N_\text{S}$ incoherent point sources are collected at $N_\text{C}$ points in the collection plane. The photons are detected only after an interferometric postprocessing.}\label{fig:scheme}
	\end{figure}
	
	Let us consider the problem of imaging $N_\text{S}$ weak, incoherent, point-like light sources with intensities $\{I_j\}_{j=1}^{N_\text{S}}$ and positions $\{\boldsymbol{r}_s=(x_s,y_s,z_s)\}_{s=1}^{N_\text{S}}$. The light emitted by the sources is collected in the collection plane, see Fig.~\ref{fig:scheme}. Conventional methods of collecting light use an aperture in the collection plane, for example a circular aperture. More generally, we can imagine that light is collected at $N_C$ points in the collection plane where arbitrary apertures can be retrieved by taking a continuous limit \cite{lupo2020quantum}. Let  $\{\boldsymbol{c}_j=(v_j, w_j)\}_{j=1}^{N_\text{C}}$ be the collection coordinates of the $N_C$ collection points.
	
	We follow the formulation of discrete quantum imaging as given by Lupo \textit{et al.} \cite{lupo2020quantum}, for a detailed derivation see Appendix \ref{app:lupo}. We assume that at most one photon is collected per collection window, known as the limit of weak sources. Further,
	we consider the paraxial regime where the distance $z_0$ of the sources from the collection plane is much larger than the source and collection coordinates, i.e., $x_s, y_s, z_s, v_j,w_j\ll z_0$, cf.~Fig.~\ref{fig:scheme}. For multiple sources, the state of a photon
	impinging on the collection plane is given by \cite{lupo2020quantum}
	\begin{align}
	\rho(\boldsymbol{p},\boldsymbol{r})=\sum_{s=1}^\Ns p_s\ket{\psi(\boldsymbol{r}_s)}\bra{\psi(\boldsymbol{r}_s)}. \label{eq:photon}
	\end{align}
	The statistical mixture in Eq.~\eqref{eq:photon} takes into account that the photon must have been emitted by one of the $N_\text{S}$ sources. The probability $p_j$ that the photon has been emitted by source $j$ is given by the relative intensity of the $j$th source, $p_j=I_j/I_{\text{tot}}$ with the total intensity $I_\text{tot}=\sum_{j=1}^{N_S}I_j$. Vectors of probabilities and source locations are defined as $\boldsymbol{p}=(p_1,\dotsc, p_{N_\text{S}})$ and $\boldsymbol{r}=(\boldsymbol{r}_1,\dotsc, \boldsymbol{r}_{N_\text{S}})$, respectively. 
	
	The photon states in the collection plane $\ket{\psi(\boldsymbol{r}_s)}$ emitted by the sources at $\boldsymbol{r}_s$ can be parameterized as
	\begin{align}
	\ket{\psi(\boldsymbol{r}_s)}=U(\boldsymbol{r}_s)\ket{\psi(0)}, \label{eq:nonortho_basis_photon}
	\end{align}
	where
	\begin{align}
	\ket{\psi(0)}=\frac{1}{\sqrt{\Nc}}\sum_{j=1}^\Nc\ket{j} \label{eq:reference}
	\end{align}
	is a reference states in the collection plane which contains the information about the location of the $\Nc$ collection points, and the unitary operator
	\begin{align}
	U(\boldsymbol{r}_s)=\e{-\ii G_xx_s-\ii G_yy_s-\ii G_zz_s} \label{eq:unitary1}
	\end{align}
	is defined via the operators
	\begin{align}
	G_x=\frac{kV}{z_0},\quad 	G_y=\frac{kW}{z_0},\quad 	G_z=\frac{k\left(V^2+W^2\right)}{2z_0^2}, \label{eq:generators}
	\end{align}
	which generate a commutative group. Here $V$ and $W$ are position operators
	in the collection plane such that $V\ket{j}=v_j\ket{j}$ and $W\ket{j}=w_j\ket{j}$ for all $j=1,\dotsc, N_S$.

	After a photon has been collected at the collection plane, it is processed coherently in a general interferometer (for details see Ref.~\cite{lupo2020quantum}) before it is measured using photodetection, cf.~Fig.~\ref{fig:scheme}. Since the interferometer corresponds to a unitary transformation which is assumed to be independent of the source parameters, it does not change the QFIM \cite{liu2019quantum} and we can proceed with calculating the QFIM from Eq.~\eqref{eq:photon}. Potential parameters of interest are the relative intensities of the sources, i.e., the probabilities $p_s$, and the positions of the sources $\boldsymbol{r}_s$, which are parameters of the unitary transformation in Eq.~\eqref{eq:unitary1}. 
	\subsection{Analytical results for discrete quantum imaging}
	Calculating the QFIM with our formal solution, Eqs.~(\ref{eq:SLDformal}-\ref{eq:qfim_nonortho}), is a matter of tedious but straightforward algebra. We choose the basis $\mathcal{B}=\left\{\ket{\psi(\boldsymbol{r}_s)}\right\}_{s=1}^{N_\text{S}}$, with basis states as given in Eq.~\eqref{eq:nonortho_basis_photon}, such that we can express the density operator \eqref{eq:photon} by a diagonal matrix $\boldsymbol{\rho}^{\mathcal{B}}(\boldsymbol{p},\boldsymbol{r})$ where $\rho_{s,s}^{\mathcal{B}}(\boldsymbol{p},\boldsymbol{r})=p_s$. If we want to estimate one of the probabilities $p_s$, for example $\theta_\mu=p_1$, the corresponding basis $\mathcal{B}_\mu$ is identical with $\mathcal{B}$ because $\partial_\mu\rho$ is supported by $\mathcal{B}$. On the other hand, if we want to estimate one of the position parameters, e.g.~$\theta_\nu=x_1$, we extend $\mathcal{B}$ by $\partial_\nu\ket{\psi(\boldsymbol{r}_s)}$ to obtain  $\mathcal{B}_\nu$. Note that $\partial_\nu\ket{\psi(\boldsymbol{r}_s)}$ is linearly independent from all vectors in $\mathcal{B}$ only for almost all values of the parameters, for more details see Appendix \ref{app:critical}.
	\subsubsection{Compatibility conditions}
	In Section \ref{sec_compatibility}, we formulated the compatibility conditions with respect to non-orthogonal bases. In the case of discrete quantum imaging, the emission properties of the sources cannot be changed and therefore the initial state is fixed for all estimation schemes. Then, the compatibility conditions reduce to (i) the commutation condition, $\Gamma_{\mu,\nu}=0$ for all $\mu\neq\nu$, where $\Gamma$ is given in Eq.~\eqref{eq:gamma},
	and (ii) the independence condition, $H_{\mu,\nu}=0$ for all $\mu\neq\nu$.
	In the following, we will refer to the commutation and independence conditions in order to interpret the results for two and three point sources.

	\subsubsection{Two sources}
	Here we consider the general problem of imaging two sources, i.e., the estimation of their positions and their relative intensity. The collected photon state is then given by
	\begin{align}
	\rho=\sum_{s=1}^2 p_s\ket{\psi(\boldsymbol{r}_s)}\bra{\psi(\boldsymbol{r}_s)}.
	\end{align}
	It is convenient to reparameterize $\rho$ using centroid and relative coordinates for the two sources:
	we define the relative coordinates as $\theta_1=\left(x_1-x_2\right)/2$, $\theta_2=\left(y_1-y_2\right)/2$, and $\theta_3=\left(z_1-z_2\right)/2$ and the centroid coordinates as $\theta_4=\left(x_1+x_2\right)/2$, $\theta_5=\left(y_1+y_2\right)/2$, and $\theta_6=\left(z_1+z_2\right)/2$.
	This means we want to estimate 7 parameters: the centroid and relative coordinates of the sources and  $\theta_7=p_1$. Note that $p_2$ is fixed due to normalization, $p_1+p_2=1$, and $p_1=I_1/I_\text{tot}$ indeed corresponds to the relative intensity. For example, if we know $I_\text{tot}$, e.g., from an independent intensity measurement, we directly obtain $I_1$ from estimating $p_1$.
	
	The matrices which must be inverted are the matrix $C$, see Eq.~\eqref{eq:c}, which is of size $2\times 2$, and the matrix $D$, see Eq.~\eqref{eq:d}, which is of size $4\times 4$.
	We find 
	\begin{align}
	H=4\begin{pmatrix}
	\operatorname{Cov}(\boldsymbol{g}) & (2p_1-1)\operatorname{Cov}(\boldsymbol{g}) & 2\left[\braket{\boldsymbol{g}(\boldsymbol{\delta}\cdot\boldsymbol{g})}-\braket{\boldsymbol{g}}\braket{\boldsymbol{\delta}\cdot\boldsymbol{g}}\right] \\
	(2p_1-1)\operatorname{Cov}(\boldsymbol{g}) & \operatorname{Cov}(\boldsymbol{g}) & 0 \\
	2\left[\braket{\boldsymbol{g}^\intercal(\boldsymbol{\delta}\cdot\boldsymbol{g})}-\braket{\boldsymbol{g}^\intercal}\braket{\boldsymbol{\delta}\cdot\boldsymbol{g}}\right]&  0 & \frac{\operatorname{Var}\left(\boldsymbol{\delta}\cdot \boldsymbol{g}\right)}{p_1(1-p_1)}
	\end{pmatrix}, \label{eq:QFIM_two_sources}
	\end{align}
	where $\boldsymbol{g}=(G_x,G_y,G_z)^\intercal$ and $\boldsymbol{\delta}=(\theta_1,\theta_2,\theta_3)^\intercal$ summarize the generators and relative source coordinates in vector notation, $\braket{\cdot}=\braket{\psi(0)|\cdot|\psi(0)}$ denotes the expectation value with respect to the reference state $\ket{\psi(0)}$, the covariance of the generators is defined as $\operatorname{Cov}(\boldsymbol{g})_{j,k}=\braket{g_jg_k}-\braket{g_j}\braket{g_k}$, and the variance as $\operatorname{Var}(A)=\braket{A^2}-\braket{A}^2$.
	
	The first two columns (rows) in Eq.~\eqref{eq:QFIM_two_sources} each consist of three columns (rows), summarized using matrix and vector notation, such that the QFIM is actually a 7-dimensional matrix. The $j$th row and column correspond to the parameter $\theta_j$.  Since we are in the paraxial regime, Eq.~\eqref{eq:QFIM_two_sources} includes only the lowest-order non-zero terms\footnote{In the full expression for the QFIM, source coordinates such as $\delta_j$ appear always as prefactors to $G_j$, where $j=x,y,z$. By factoring out $1/z_0$ from $G_j$, we obtain rescaled source coordinates such as $\delta_j'=\delta_j/z_0$. We expanded the coefficients of the QFIM with respect to these rescaled sources coordinates up the first non-vanishing order. This yields Eq.~\eqref{eq:QFIM_two_sources}.} in the rescaled source coordinates $\delta_j/z_0$.

	By fixing certain parameters, we recover known results: if we set $\theta_7=p_1=1/2$, i.e., we consider sources of equal intensity, the upper-left $6\times 6$ block corresponding to the centroid and relative coordinates becomes block diagonal which reproduces the well-known result that, according to the independence condition, centroid and relative coordinates can be estimated independently from each other. On the other hand, the estimation errors among different centroid coordinates are not independent (the same holds for different relative coordinates) \cite{prasad2019quantum, napoli2019towards, lupo2020quantum}. Further, from the zero blocks in Eq.~\eqref{eq:QFIM_two_sources} it follows that the relative coordinates can be estimated independently of the relative intensity of the sources. This is in agreement with the results for two sources constrained to one dimension \cite{rehacek2017multiparameter}. Finally, Eq.~\eqref{eq:QFIM_two_sources} generalizes the result for the estimation of centroid and relative coordinates of two sources of known unequal brightness \cite{prasad2020quantum}.  We also reproduce the result of Ref.~\cite{prasad2020quantum} that for unequal source brightness the estimation of centroid and relative coordinates is no longer independent; in particular, since the upper-left 6$\times$6 block is proportional to $\operatorname{Cov}(\boldsymbol{g})$, there exists no collection pattern or aperture which makes the off-diagonal vanish while the diagonal blocks are nonzero.
	
	We note that the QFIM is independent of the centroid coordinates, i.e., the information content of the collected light does not change by jointly moving the sources or the collection instrument as a whole. Further, the covariance $\operatorname{Cov}(\boldsymbol{g})$ which characterizes the QFIM for relative and centroid coordinates shows that distributions of collection points with a large variance along a $x$-axis ($y$-axis) are better suited to estimate the corresponding source coordinates along the $x$ ($y$) direction, while the variance along the $x$- and $y$-axis contributes equally for the estimation of source coordinates along the $z$ direction. Similarly, a better precision in estimating the relative intensity can be achieved if the collection points exhibit a larger variance along the $x$-axis ($y$-axis) if the sources have a larger separation along the $x$-axis ($y$-axis) compared to the $y$-axis ($x$-axis), because the variances in $H_{7,7}$ are scaled with the squared sources separation along the corresponding direction; if the sources are only separated along the $z$-axis, the variances of the source coordinates along the $x$- and $y$-axis contribute equally. For example, for a circular aperture of diameter $D$, the variance along the $x$- or $y$-axis is proportional to $D^2$, i.e., the diagonal elements of the QFIM scale as $D^2$.

	To answer the question of the existence of optimal measurements, we compute the matrix $\boldsymbol{\Gamma}$. Using the same block-partitioning as for the QFIM in Eq.~\eqref{eq:QFIM_two_sources}, we find
	\begin{align}
	\boldsymbol{\Gamma}=4\begin{pmatrix}
	0 & \boldsymbol{\Gamma}_{12} & \boldsymbol{\Gamma}_{13} \\
	-\boldsymbol{\Gamma}_{12}^\intercal & 0 & \boldsymbol{\Gamma}_{23} \\
	-\boldsymbol{\Gamma}_{13}^\intercal & -\boldsymbol{\Gamma}_{23}^\intercal & 0
	\end{pmatrix}, \label{eq:gamma_two_sources}
	\end{align}
	where
	\begin{align}
	\boldsymbol{\Gamma}_{23}&= 2\left[\braket{\boldsymbol{g}\left(\boldsymbol{\delta}\cdot\boldsymbol{g}\right)^2}-\braket{\boldsymbol{g}}\braket{\left(\boldsymbol{\delta}\cdot\boldsymbol{g}\right)^2}+2\braket{\boldsymbol{g}}\braket{\boldsymbol{\delta}\cdot\boldsymbol{g}}^2-2\braket{\boldsymbol{g}\left(\boldsymbol{\delta}\cdot\boldsymbol{g}\right)}\braket{\boldsymbol{\delta}\cdot\boldsymbol{g}}\right]\,,\\
	\boldsymbol{\Gamma}_{13}&= \left(2p_1-1\right)\boldsymbol{\Gamma}_{23}\,,\\
	\boldsymbol{\Gamma}_{12}&=2p_1(p_1-1)\boldsymbol{\Gamma}_{23}\left[\braket{\boldsymbol{g}^\intercal\left(\boldsymbol{\delta}\cdot\boldsymbol{g}\right)}-\braket{\boldsymbol{g}^\intercal}\braket{\boldsymbol{\delta}\cdot\boldsymbol{g}}\right]/\operatorname{Var}\left(\boldsymbol{\delta}\cdot \boldsymbol{g}\right)\,.
	\end{align}
	Similarly to the lowest-order approximation of the QFIM, the coefficients of $\boldsymbol{\Gamma}$ are given at the lowest non-vanishing order with respect to the rescaled source coordinates $\delta_j/z_0$. 
	We find that the diagonal blocks of $\boldsymbol{\Gamma}$ vanish. Thus, there exist optimal measurements for the estimation of centroid or relative coordinates (but not for the combination of both) and the QCRB can be saturated although the estimators for different centroid or relative coordinates are generally correlated since the QFIM is not diagonal. Further, we note that $\boldsymbol{\Gamma}_{23}$ and $\boldsymbol{\Gamma}_{13}$ are second-order with respect to $\boldsymbol{\delta}/z_0$ while $\boldsymbol{\Gamma}_{12}$ is first order. This means that in the the paraxial regime  $\boldsymbol{\Gamma}_{23}$ and $\boldsymbol{\Gamma}_{13}$ are approximately zero and there exist optimal measurements for the relative intensity and the relative coordinates or the relative intensity and the centroid coordinates.  We can infer from the zeros in Eq.~\eqref{eq:QFIM_two_sources} that one centroid coordinate and the relative intensity can be jointly estimated optimally, i.e., with independent estimators and measurements which saturate the QCRB.
	
	Important examples of apertures are the circular aperture, e.g., studied in Refs.~\cite{yu2018quantum, prasad2019quantum, prasad2020quantum}, and the Gaussian-beam assumption \cite{napoli2019towards, bisketzi2019quantum} which corresponds to a Gaussian distribution of collection points. Since we found a general solution for any aperture or distribution of collection points, we can make more general statements. For example, for any distribution of collection points which is radially symmetric around the optical axis, i.e., such that for any collection point at position $(v,w)$ there exists another collection point at $(-v,-w)$, we find that Eq.~\eqref{eq:gamma_two_sources} becomes $\boldsymbol{\Gamma}=0$, i.e., the QCRB can in principle be saturated. Clearly, this includes the aforementioned cases of circular aperture or a Gaussian distribution of collection points.

	\subsubsection{Three sources}
	We further consider two examples with three equidistant sources. We assume that the three sources are aligned along the $x$-axis, i.e., $x_1=c_x-\delta_x$, $x_2=c_x$, and $x_3=c_x+\delta_x$, with known centroid coordinate $c_x$. 
	First, let us consider the problem of estimating the distance $\delta_x$ where we assume that the relative intensities are known. In lowest order with respect to $\delta_x/z_0$, the QFI is then obtained as
	\begin{align}
	H=4\left(1-p_2\right)\operatorname{Var}\left(g_{x}\right), \label{eq:3sources_distance}
	\end{align}
	where $p_1+p_2+p_3=1$. In accordance with our results for two sources, we find that a non-zero variance in the collection points in the $x$ direction is crucial for the estimation of the relative intensity if the sources are aligned along the $x$-axis. 
	For comparison, in the distance estimation of two sources of known (unequal) relative intensity along the $x$-axis, i.e., $x_1=c_x-\delta_x/2$, and $x_2=c_x+\delta_x/2$, with known centroid coordinate $c_x$, one finds 
	$H=\operatorname{Var}\left(g_{x}\right)$ for the QFI. At first sight, this is surprising because it does not seem to match Eq.~\eqref{eq:3sources_distance} which yields $H=4p_3\operatorname{Var}\left(g_{x}\right)$ if we set $p_1$ to zero. However, the two estimation problems are different: in case (i), the three-sources problem with $p_1=0$  corresponds to a situation where the position of source 2 is known, while the position of source 3 has to be estimated (if $\delta_x$ is varied, only source 3 moves). Accordingly, the QFI equals the QFI for the position estimation of a single source rescaled with its relative intensity $p_3$; source two effectively acts as a source of noise which reduces the relevant signal. On the other hand, in case (ii), the problem of the two sources placed symmetrically around a known centroid coordinate requires a joint estimation of the sources in order to estimate their distance (if $\delta_x$ is varied, both source move).
	
	Both cases can be summarized in one formula by parameterizing the two sources as $x_1=c_x+\delta_x (q-1/2)$, and $x_2=c_x+\delta_x (q+1/2)$ where $q\in \mathbb{R}$ is an additional scaling parameter such that we obtain case (i) for $q=1/2$ and case (ii) for $q=0$. Then, we find for the QFI in lowest order with respect to $\delta_x/z_0$,
	\begin{align}
	H=\left[1+4q^2+4q\left(2p_2-1\right)\right]\operatorname{Var}\left(g_{x}\right). \label{eq:2sources_distance}
	\end{align} 
	Note that $H$ grows quadratically with $q$, a typical effect when parameters are rescaled because scaling factors increase the sensitivity with respect to changes of the parameters. Finally, it is worth noticing that we can find analog results for the QFI in Eqs.~\eqref{eq:3sources_distance} and \eqref{eq:2sources_distance} if the sources are aligned along the $y$-axis ($z$-axis) where $\operatorname{Var}\left(g_{x}\right)$ is replaced with $\operatorname{Var}\left(g_{y}\right)$ ($\operatorname{Var}\left[g_{z}\right]$).

	For our next example, we consider again three equidistant sources of unequal brightness along the $x$-axis, $x_1=c_x-\delta_x$, $x_2=c_x$, and $x_3=c_x+\delta_x$, with known centroid coordinate. This time, we assume that their distance is known and we estimate $ p_{1} $ and $ p_{2} $ while $p_3$ is determined by $ p_{3}=1-p_{1}-p_{2} $. In lowest order with respect to $\delta_x/z_0$, the QFIM is found to be
	\begin{align}
	\boldsymbol{H}\left(p_1,p_2\right)=\frac{\delta_x^{2}\operatorname{Var}\left(g_{x}\right)}{\left(1-p_{2}\right)\left(4p_{1}+p_{2}\right) -4p_{1}^{2}}\begin{pmatrix}
	16\left(1-p_2\right) & 4 \left(1+2p_1-p_2\right) \\
	4 \left(1+2p_1-p_2\right) & \left(1+8p_{1}\right)
	\end{pmatrix}.\label{eq:qfim_3intensity}
	\end{align}
	The QFIM is proportional to $\delta_x^2 \operatorname{Var}\left(g_{x}\right)$, which means that a nonzero source separation as well as a nonzero separation of collection points along the $x$-axis is necessary to estimate the relative intensities. The fact that source 2 has a special place as the middle source breaks the symmetry between source 1 and 2. This is reflected in the asymmetry between the diagonal coefficients of the QFIM which correspond to the estimation of $p_1$ and $p_2$. For example, if we evaluate the QFIM for equal source intensities, $p_1=p_2=1/3$ (i.e., $p_3=1/3$), the asymmetry persists:
	\begin{align}
	\boldsymbol{H}\left(\frac{1}{3},\frac{1}{3}\right)=\delta_x^{2}\operatorname{Var}\left(g_{x}\right)\begin{pmatrix}
	16 & 8 \\
	8 & \frac{11}{2}
	\end{pmatrix}. \label{eq:qfim_3intensity1/3}
	\end{align}
	Further, note that there are statistical correlations between the estimators of $p_1$ and $p_2$ because the off-diagonal coefficients in Eq.~\eqref{eq:qfim_3intensity} are nonzero for sources of finite brightness. We find that $\boldsymbol{\Gamma}$ is zero up to second order with respect to the rescaled source coordinates (while there are non-zero higher-order terms). This means that, in the paraxial regime, there exist optimal measurements for the joint estimation of the relative intensities.

	\section{Discussion and conclusion} \label{sec:discussion}
	In this paper we obtained fully general solutions for the quantum Fisher information (QFI) and the QFI matrix (QFIM) [see Eq.~\eqref{eq:SLDformal} and \eqref{eq:qfim_nonortho}], which are figure of merits of core importance in  quantum metrology. Based on these solutions, we provided a general method to analytically calculate the QFI and QFIM. 
	Our solutions generalize previous results \cite{genoni2019non,bisketzi2019quantum} and we show in particular that \v{S}afr\'anek's QFIM expression \cite{safranek2018simple} is a special case of our solution for orthogonal bases.
	
	Compared to the conventional method of calculating the QFIM which relies on matrix diagonalization, our method does not share the rather strict limitations of analytical matrix diagonalization. 
	Instead, our method relies on the computation of matrix inverses which amounts to solving linear systems. Then, finding analytical solutions is usually only limited by our ability to handle long algebraic expressions, see Appendix \ref{app:analytics} for more details.

	While \v{S}afr\'anek's QFIM expression \cite{safranek2018simple} shares the aforementioned advantages from avoiding matrix diagonalization, our method has the additional advantage that it does not rely on expanding operators in an orthogonal basis. This can be very convenient when the density operator is given in a non-orthogonal basis. In such cases good choices for an orthogonal basis for analytical computations are often hard to find and lead to an inefficient representation which hampers analytical computation of the QFIM. In particular, switching to an orthogonal basis often leads to larger matrices which do not have full rank. Our method accepts any density matrix with respect to any given, possibly non-orthogonal basis where the only requirement is that the density matrix has full rank in this basis. Then, our method keeps the dimension of matrices as low as possible which facilitates analytical computations significantly. While our method avoids diagonalization, it requires the computation of two inverse matrices of dimension $d$ and $d^2$ where $d$ is the rank of the density matrix. Since matrix inversion is simpler than diagonalization in many respects, our method should be applied whenever it is impossible to diagonalize the density matrix. We remark once more that our method and the above considerations apply to the analytical evaluation of the QFIM. A detailed discussion about numerical computation is provided in Appendix \ref{app:numerics}.

	We demonstrated the usefulness of our method by deriving  new analytical solutions for discrete quantum imaging which generalize existing results for two points sources and provide insights about selected problems with three point sources. Thanks to its generality and its advantages over previous methods when the density operator is given in a non-orthogonal basis, we expect that our method of computing analytical solutions for the QFI and the QFIM will find widespread application and can become a standard tool in quantum metrology.

	\appendix

	\section{Derivation of the QFIM for general bases} \label{app:derivation}
	In this appendix, we will derive general expressions for the QFIM. The derivation involves four steps: (i) we write the Lyapunov equation for matrices defined with respect to an arbitrary (non-orthogonal) basis, (ii) we rewrite the Lyapunov equations using block-vectorization,  (iii) we derive a formal solution for the SLD, and (iv) we insert the solution in the expression for the QFIM. 
	
	We would like to refer to the main text [from the beginning of Section \ref{sec:general_result} up to but not including Eq.~\eqref{eq:SLDformal}] for Lyapunov equations expressed in a general (non-orthogonal) basis and some notation which will be used in the following. We continue with deriving a formal solution of the Lyapunov equations. We start by introducing more notation. For a given block decomposition of matrix $ \boldsymbol{A} $, $\operatorname{vecb}(\boldsymbol{A})$ denotes a block-wise vectorization: 
	\begin{align}
	\boldsymbol{A}=\begin{bmatrix}
	\boldsymbol{A}_{11} & \boldsymbol{A}_{12} \\
	\boldsymbol{A}_{21} & \boldsymbol{A}_{22}
	\end{bmatrix}, \quad
	\operatorname{vecb}(\boldsymbol{A}):=\begin{pmatrix}
	\operatorname{vec}\left(\boldsymbol{A}_{11}\right) \\
	\operatorname{vec}\left(\boldsymbol{A}_{21}\right) \\
	\operatorname{vec}\left(\boldsymbol{A}_{12}\right) \\
	\operatorname{vec}\left(\boldsymbol{A}_{22}\right)
	\end{pmatrix}. \label{eq:app:vecb}
	\end{align}
	Note that $ \operatorname{vecb} $ depends on the particular partitioning of matrix $ A $. We use the notation that $\boldsymbol{A}_{ij}$ denotes the $(i,j)$th block of $\boldsymbol{A}$ in contrast to $A_{i,j}$ which denotes the $(i,j)$th coefficients of matrix $\boldsymbol{A}$.  
	In particular, we apply a convention for dividing matrices in blocks which has been introduced in the main text (Section \ref{sec:general_result}).
	
	Next we rewrite the SLD equations using an important identity for block-vectorization \cite[p.49]{singh1972some}, see also Ref.~\cite[Lemma 4]{tracy1989partitioned}:
	\begin{align}
	\vecb\left(\boldsymbol{ABC}\right)=(\boldsymbol{C}^\intercal\odot \boldsymbol{A}) \vecb\left(\boldsymbol{B}\right),\label{eq:identity}
	\end{align}
	where $\boldsymbol{C}^\intercal$ denotes the transpose of matrix $\boldsymbol{C}$, and $\odot$ denotes the Tracy--Singh product, a generalization of the tensor product for block-partitioned matrices. For two partitioned matrices $\boldsymbol{A}$ and $\boldsymbol{B}$, it is defined as
	\begin{align}
	\boldsymbol{A}\odot \boldsymbol{B}=\begin{pmatrix}
	\boldsymbol{A}_{11}\otimes \boldsymbol{B}_{11} & \boldsymbol{A}_{11}\otimes \boldsymbol{B}_{12} & \boldsymbol{A}_{12}\otimes \boldsymbol{B}_{11} & \boldsymbol{A}_{12}\otimes \boldsymbol{B}_{12} \\
	\boldsymbol{A}_{11}\otimes \boldsymbol{B}_{21} & \boldsymbol{A}_{11}\otimes \boldsymbol{B}_{22} & \boldsymbol{A}_{12}\otimes \boldsymbol{B}_{21} & \boldsymbol{A}_{12}\otimes \boldsymbol{B}_{22} \\
	\boldsymbol{A}_{21}\otimes \boldsymbol{B}_{11} & \boldsymbol{A}_{21}\otimes \boldsymbol{B}_{12} & \boldsymbol{A}_{22}\otimes \boldsymbol{B}_{11} & \boldsymbol{A}_{22}\otimes \boldsymbol{B}_{12} \\
	\boldsymbol{A}_{21}\otimes \boldsymbol{B}_{21} & \boldsymbol{A}_{21}\otimes \boldsymbol{B}_{22} & \boldsymbol{A}_{22}\otimes \boldsymbol{B}_{21} & \boldsymbol{A}_{22}\otimes \boldsymbol{B}_{22} 
	\end{pmatrix}.
	\end{align}
	An important special case of Eq.~\eqref{eq:identity} is $\vecc\left(\boldsymbol{ABC}\right)=(\boldsymbol{C}^\intercal\otimes \boldsymbol{A}) \vecc\left(\boldsymbol{B}\right)$.	
	In block-vectorized from, the Lyapunov equation for $\boldsymbol{L}^{\mathcal{B}_{\mu}}_\mu$ reads
	\begin{align}
	2\vecb\left(\boldsymbol{\partial_\mu\rho}\right) &=\vecb\left(\boldsymbol{L}_\mu \boldsymbol{G}  \boldsymbol{\rho} +\boldsymbol{\rho}  \boldsymbol{G}  \boldsymbol{L}_\mu\right).\label{eq:app:lyapunov_block}
	\end{align}
	All matrices in Eq.~\eqref{eq:app:lyapunov_block} are give with respect to the basis $\mathcal{B}_{\mu}$, however, for better readability we drop the superscripts  $\mathcal{B}_{\mu}$ in Eq.~\eqref{eq:app:lyapunov_block} in the following until we reach Eq.~\eqref{eq:app:solution22}.  
	With the identity \eqref{eq:identity}, and using that $\boldsymbol{\rho}$ and $\boldsymbol{G}$ are hermitian, we find
	\begin{align}
	2\vecb\left(\boldsymbol{\partial_\mu\rho}\right)&=\vecb\left(\boldsymbol{L}_\mu \boldsymbol{G} \boldsymbol{\rho} \right)+\vecb\left(\boldsymbol{\rho}\boldsymbol{G} \boldsymbol{L}_\mu\right)\\
	&=\vecb\left(\one  \boldsymbol{L}_\mu \boldsymbol{G} \boldsymbol{\rho} \right)+\vecb\left(\boldsymbol{\rho}\boldsymbol{G} \boldsymbol{L}_\mu \one \right) \label{eq:add_1}\\
	&=\left(\boldsymbol{G} \boldsymbol{\rho} \right)^\intercal\odot\one \vecb\left(\boldsymbol{L}_\mu \right)+\one \odot \left(\boldsymbol{\rho} \boldsymbol{G}\right)\vecb\left( \boldsymbol{L}_\mu \right) \label{eq:use_identity}\\
	&=\overline{ \boldsymbol{\rho} \boldsymbol{G}}\odot\one \vecb\left(\boldsymbol{L}_\mu \right)+\one \odot \left(\boldsymbol{\rho} \boldsymbol{G}\right)\vecb\left( \boldsymbol{L}_\mu \right) \\
	&=\left[\overline{ \boldsymbol{\rho} \boldsymbol{G}}\odot\one +\one \odot \left(\boldsymbol{\rho} \boldsymbol{G}\right)\right]\vecb\left( \boldsymbol{L}_\mu \right),  \label{eq:sld_vecb}
	\end{align}
	where $\overline{\boldsymbol{A}}$ denotes the complex conjugate of $\boldsymbol{A}$.
	By decomposing $\boldsymbol{\rho}$ and $\boldsymbol{G}$ in blocks as stated below Eq.~\eqref{eq:rhoblock}, Eq.~\eqref{eq:sld_vecb} becomes (still, all matrices with respect to $\mathcal{B}_{\mu}$)
	\begin{align}
	2\begin{pmatrix}
	\vecc\left[\left(\boldsymbol{\partial_\mu\rho}\right) _{11}\right]\\
	\vecc\left[\left(\boldsymbol{\partial_\mu\rho}\right) _{21}\right]\\
	\vecc\left[\left(\boldsymbol{\partial_\mu\rho}\right) _{12}\right]\\
	0
	\end{pmatrix}&=\begin{pmatrix}\one_{11}\otimes  \left(\boldsymbol{\rho}_{11} \boldsymbol{G}_{11}\right) + \overline{ \boldsymbol{\rho}_{11} \boldsymbol{G}_{11}}\otimes \one_{11} & \one_{11}\otimes  \left(\boldsymbol{\rho}_{11} \boldsymbol{G}_{12}\right) & \overline{ \boldsymbol{\rho}_{11} \boldsymbol{G}_{12}}\otimes \one_{11} & 0\\
	0 & \overline{ \boldsymbol{\rho}_{11} \boldsymbol{G}_{11}}\otimes\one_{22} & 0 & \overline{ \boldsymbol{\rho}_{11} \boldsymbol{G}_{12}}\otimes \one_{22} \\
	0 & 0 & \one_{22}\otimes  \left(\boldsymbol{\rho}_{11} \boldsymbol{G}_{11}\right) & \one_{22}\otimes  \left(\boldsymbol{\rho}_{11} \boldsymbol{G}_{12}\right)\\
	0 & 0 & 0 & 0 
	\end{pmatrix}\begin{pmatrix}
	\vecc\left[\left(\boldsymbol{L}_\mu\right)_{11}\right]\\
	\vecc\left[\left(\boldsymbol{L}_\mu\right)_{21}\right]\\
	\vecc\left[\left(\boldsymbol{L}_\mu\right)_{12}\right]\\
	\vecc\left[\left(\boldsymbol{L}_\mu\right)_{22}\right]
	\end{pmatrix}, \label{eq:app:big_one}
	\end{align}
	where it followed from the chain rule of differentiation that $\vecc\left[\left(\partial_\mu\rho\right) _{22}\right]$ is zero.
	
	We continue by solving Eq.~\eqref{eq:app:big_one} for the SLD $L_\mu$. 
	Note that the big matrix of block matrices in Eq.~\eqref{eq:app:big_one} is already  upper triangular, and the bottom row contains only zero blocks indicating that $\boldsymbol{L}_\mu$ is underdetermined. We pick a solution for $\boldsymbol{L}_\mu$ by setting $\vecc \left[\left(\boldsymbol{L}_{\mu}\right)_{22}\right]=0$. This yields the following solution for the SLD:
	\begin{align}
	\vecc \left[\left(\boldsymbol{L}_{\mu}\right)_{11}\right]&=2\,\boldsymbol{D}^{-1}\vecc\left[\left(\boldsymbol{\partial_\mu\rho}\right) _{11}-\boldsymbol{E}-\boldsymbol{E}^\dagger\right],\label{eq:app:solution11}\\
	\vecc \left[\left(\boldsymbol{L}_{\mu}\right)_{21}\right]&=2\,\vecc\left[\left(\boldsymbol{\partial_\mu\rho}\right) _{21}\left(\boldsymbol{C}^{-1}\right)^\dagger\right],\\
	\vecc \left[\left(\boldsymbol{L}_{\mu}\right)_{12}\right]&=2\,\vecc\left[\boldsymbol{C}^{-1}\left(\boldsymbol{\partial_\mu\rho}\right) _{12}\right],  \label{eq:app:solution12}\\
	\vecc \left[\left(\boldsymbol{L}_{\mu}\right)_{22}\right]&=0, \label{eq:app:solution22}
	\end{align}
	where we used identity \eqref{eq:identity} for vectorization, and the matrices $\boldsymbol{C}$, $\boldsymbol{D}$, and $\boldsymbol{E}$ are defined in the main text, see Eqs.~(\ref{eq:c}-\ref{eq:e}).

	Eqs.~(\ref{eq:app:solution11}-\ref{eq:app:solution22}) constitute a general solution to the Lyapunov equations \eqref{eq:sld_eqs} with respect to the basis $\mathcal{B}_{\mu}$. In particular, using identity \eqref{eq:identity} and the $\operatorname{mat}(\cdot)$ operation [defined in the main text, cf.~Eq.~\eqref{eq:mat_example}], we can write the solution for $L_\mu$ in matrix form where we keep in mind that all matrices starting from Eq.~\eqref{eq:app:lyapunov_block} up to the following equation are given with respect to $\mathcal{B}_\mu$, though, we dropped the superscript $\mathcal{B}_\mu$ for better readability. The solution of $\boldsymbol{L}_\mu$ in matrix form is given in Eq.~\eqref{eq:SLDformal} in the main text where we also explain how to obtain the QFIM [Eq.~\eqref{eq:qfim_nonortho}].
	
	An alternative representation of the QFIM to Eq.~\eqref{eq:qfim_nonortho} can be obtained by writing the QFIM in vectorized form,
	\begin{align}
	H_{\mu,\nu}&=\tr\left[\boldsymbol{L}^{\mathcal{B}_{\mu,\nu}}_{\mu} \boldsymbol{G}^{\mathcal{B}_{\mu,\nu}}(\boldsymbol{\partial_\nu\rho})^{\mathcal{B}_{\mu,\nu}}\boldsymbol{G}^{\mathcal{B}_{\mu,\nu}}\right]\\
	&=\tr\left[\boldsymbol{G}^{\mathcal{B}_{\mu,\nu}}(\boldsymbol{\partial_\nu\rho})^{\mathcal{B}_{\mu,\nu}} \boldsymbol{G}^{\mathcal{B}_{\mu,\nu}}\boldsymbol{L}^{\mathcal{B}_{\mu,\nu}}_{\mu}\right]\\
	&=\vecb\left[\boldsymbol{G}^{\mathcal{B}_{\mu,\nu}}(\boldsymbol{\partial_\nu\rho})^{\mathcal{B}_{\mu,\nu}} \boldsymbol{G}^{\mathcal{B}_{\mu,\nu}}\right]^\dagger\vecb\left(\boldsymbol{L}^{\mathcal{B}_{\mu,\nu}}_{\mu}\right), \label{eq:QFIM_blockvec}
	\end{align}
	where we used the cyclic property of the trace, and $\trb{\boldsymbol{AB}}=\vecb\left(\boldsymbol{A}\right)^\dagger\vecb\left(\boldsymbol{B}\right)$. The block-vectorized solution Eqs.~(\ref{eq:app:solution11}-\ref{eq:app:solution22}) can be extended to the basis $\mathcal{B}_{\mu,\nu}$ by padding with zeros and can then be plugged into Eq.~\eqref{eq:QFIM_blockvec}.
	\section{A quantum formulation of discrete quantum imaging} \label{app:lupo}
	In this appendix, we will reproduce the formulation of discrete quantum imaging as given by Lupo \textit{et al.} \cite{lupo2020quantum}. We assume that at most one photon is collected per collection window, known as the limit of weak sources. Then, 
	a general single-mode photon state emitted by a source at $\boldsymbol{r}_s$ and collected in the collection plane can be expressed as
	\begin{align}
	\ket{\psi(\boldsymbol{r}_s)}=\sum_{j=1}^{N_\text{C}}\gamma\left(\boldsymbol{c}_j,\boldsymbol{r}_s\right)\ket{j}.
	\end{align}
	The orthonormal states $\ket{j}$ contain the information about the location of the collection points where $j$ labels the collection points, and $\gamma\left(\boldsymbol{c}_j,\boldsymbol{r}_s\right)$ are general complex amplitudes which must fulfill normalization, $\sum_{j}\left|\gamma\left(\boldsymbol{c}_j,\boldsymbol{r}_s\right)\right|^2=1$.
	The relative phases
	\begin{align}
	\varphi\left(\boldsymbol{c}_j,\boldsymbol{r}_s\right)=\operatorname{arg}\gamma\left(\boldsymbol{c}_j,\boldsymbol{r}_s\right)= k l\left(\boldsymbol{c}_j,\boldsymbol{r}_s\right)
	\end{align}
	depend on the wave number $k$ and the path length $l\left(\boldsymbol{c}_j,\boldsymbol{r}_s\right)$ from the source at $\boldsymbol{r}_s$ to the collection point $\boldsymbol{c}_j$,
	\begin{align}
	l\left(\boldsymbol{c}_j,\boldsymbol{r}_s\right)&=\sqrt{\left(x_s-v_j\right)^2+\left(y_s-w_j\right)^2+\left(z_s+z_0\right)^2}\\
	&=z_0\sqrt{\left(x'_s-v'_j\right)^2+\left(y'_s-w'_j\right)^2+\left(z'_s+1\right)^2}, \label{eq:app:second_line}
	\end{align}
	where, in the second line \eqref{eq:app:second_line}, the distance $z_0$ of the sources from the collection plane, cf.~Fig.~\ref{fig:scheme}, has been factored out and the primed variables equal the unprimed ones scaled with a factor $1/z_0$.

	We consider the paraxial regime where the distance of the sources from the collection plane is much larger than the source and collection coordinates, i.e., $x_s, y_s, z_s, v_j,w_j\ll z_0$; this means we can expand $\varphi\left(\boldsymbol{c}_j,\boldsymbol{r}_s\right)$ for all primed variables around zero. In order to get non-trivial terms for each of the source coordinates, we compute a multivariate Taylor expansion up to the third order:
	\begin{align}
	\varphi\left(\boldsymbol{c}_j,\boldsymbol{r}_s\right)\simeq k z_0\left[\left(z'_s-1\right)\left(x'_sv'_j+y'_sw'_j\right)-z'_s\frac{{v'_j}^2+{w'_j}^2}{2}+1+z'_s+\frac{\left({x'_s}^2+{y'_s}^2\right)\left(1-z'_s\right)+{v'_j}^2+{w'_j}^2}{2}\right]. \label{eq:app:phase3}
	\end{align}
	Note that we can drop terms which depend only on the collection plane coordinates (and not on the source coordinates) because those terms can be absorbed in the definition of $\ket{j}$, and the phases which depend only on the source coordinates correspond to trivial phase factors which are canceled out when writing down the density operator [cf.~Eq.~\eqref{eq:photon}]. Then, we find
	\begin{align}
	\varphi\left(\boldsymbol{c}_j,\boldsymbol{r}_s\right)\simeq k z_0\left[\left(z'_s-1\right)\left(x'_sv'_j+y'_sw'_j\right)-z'_s\frac{{v'_j}^2+{w'_j}^2}{2}\right]. \label{eq:app:phase}
	\end{align}
	Note that the only non-trivial $z'_s$ term in Eq.~\eqref{eq:app:phase} is of third order with respect to the variables $z'_s, v'_j$, and $w'_j$, which is the reason why we had to compute the multivariate Taylor expansion up to the third order.
	
	Since $x_s, y_s, z_s, v_j,w_j\ll z_0$ and since the sources are incoherent, we can assume that a photon is detected with equal probability at one of the collection points, such that
	\begin{align}
	\gamma\left(\boldsymbol{c}_j,\boldsymbol{r}_s\right)=\frac{1}{\sqrt{N_\text{C}}}\e{\ii\varphi\left(\boldsymbol{c}_j,\boldsymbol{r}_s\right)}.
	\end{align}
	By introducing position operators
	$V$ and $W$ in the collection plane such that $V\ket{j}=v_j\ket{j}$ and $W\ket{j}=w_j\ket{j}$ for all $j$, we can define a unitary operator which generates the relative phases by acting on a reference state $\ket{\psi(0)}$ which does not depend on the source coordinates:
	\begin{align}
	\ket{\psi(\boldsymbol{r}_s)}=U(\boldsymbol{r}_s)\ket{\psi(0)}, \label{eq:app:nonortho_basis_photon}
	\end{align}
	where
	\begin{align}
	\ket{\psi(0)}=\frac{1}{\sqrt{\Nc}}\sum_{j=1}^\Nc\ket{j}. \label{eq:app:reference}
	\end{align}
	By replacing the collection plane coordinates in Eq.~\eqref{eq:app:phase} with the position operators $V$ and $W$, and keeping only first-order terms in the source coordinates\footnote{It would also be possible to keep all phase terms in Eq.~\eqref{eq:app:phase}, however, restricting ourselves to a first-order approximation in the source coordinates is justified in the paraxial regime and it simplifies our calculations.}, we obtain the unitary operator
	\begin{align}
	U(\boldsymbol{r}_s)=\e{-\ii G_xx_s-\ii G_yy_s-\ii G_zz_s}, \label{eq:app:unitary1}
	\end{align}
	where we defined the operators
	\begin{align}
	G_x=\frac{kV}{z_0},\quad 	G_y=\frac{kW}{z_0},\quad 	G_z=\frac{k\left(V^2+W^2\right)}{2z_0^2}, \label{eq:app:generators}
	\end{align}
	as generators of a commutative group.
	
	For multiple sources, the state of a photon
	impinging on the collection plane is given by
	\begin{align}
	\rho(\boldsymbol{p},\boldsymbol{r})=\sum_{s=1}^\Ns p_s\ket{\psi(\boldsymbol{r}_s)}\bra{\psi(\boldsymbol{r}_s)}. \label{eq:app:photon}
	\end{align}
	The statistical mixture in Eq.~\eqref{eq:app:photon} takes into account that the photon must have been emitted by one of the $N_\text{S}$ incoherent sources. The probability $p_s$ that the photon has been emitted by source $j$ is given by the relative intensity of the $j$th source, $p_j=I_j/I_{\text{tot}}$ with the total intensity $I_\text{tot}=\sum_{j=1}^{N_S}I_j$. Vectors of probabilities and source locations are defined as $\boldsymbol{p}=(p_1,\dotsc, p_{N_\text{S}})$ and $\boldsymbol{r}=(\boldsymbol{r}_1,\dotsc, \boldsymbol{r}_{N_\text{S}})$. 
	\section{Critical points and the reduction of the QFIM to lower order terms} \label{app:critical}
	Our approach of calculating the QFIM makes use of non-orthogonal bases. By definition, this implies that all basis vectors are linearly independent. However, since the basis vectors can depend on the parameters of interest, it can happen that for special values of those parameters, which we call critical points, the set of basis vectors is no longer linearly independent and, thus, it does no longer constitute a valid basis. This typically happens when we choose the parameters such that the dimensionality of the problem reduces; for example, when setting the distance of two sources to zero, the two sources effectively become one source. For such critical points, our method can still be applied but a smaller basis has to be used. Typically, the coefficients of the QFIM are well-defined and smooth functions with respect to the parameters. Only at the critical points, we find removable singularities which can be eliminated by separately calculating the QFIM at the critical points using a reduced set of basis vectors.
	
	When we try to reduce a higher order expression for the QFIM in Section \ref{sec:discrete_q_imaging} to first or second order, the problem of critical points becomes relevant if we try to expand around such points. In order to calculate a Taylor expansion around such removable singularities, we take the limit towards the singularity. The limit can be calculated making repeatedly use of L'H\^opital's rule. The only drawback of this method is that iterating L'H\^opital's rule requires to take higher orders of derivatives which leads to even longer expressions. In some cases, this turned out to be a computational bottleneck.
	
	\section{Numerical computation of the QFIM}\label{app:numerics}

	Our main focus in this work has been on the  analytical computation of the QFIM. At the same time, the introduction of a new expression for the QFIM, as given in Eqs.~\eqref{eq:SLDformal} and \eqref{eq:qfim_nonortho}, might raise questions regarding its usefulness for numerical computations. However, the challenges of analytical and numerical computation are very different.

	On the one hand, as we discussed in Section \ref{sec:prelim}, analytical matrix diagonalization is notoriously hard and usually impossible for matrices of rank larger than $4$. On the other hand, when it comes to numerics, it has been shown that most standard linear algebra operations, including matrix diagonalization, matrix inversion, and solving Lyapunov equations, can be done numerically in a stable way and asymptotically as fast as matrix multiplication \cite{demmel2007fast}. The complexity of matrix multiplication is not known. Naive algorithms multiply two $d\times d$ matrix in approximately $\mathcal{O}(d^{3})$ time. From all known algorithms the one with the best asymptotic complexity takes approximately $\mathcal{O}(d^{2.37})$ time \cite{alman2021refined}. However, due to a large constant factor hidden by the $\mathcal{O}$ notation, this and other algorithms with a similar complexity are not practical.
	
	This means that in terms of computational complexity there is no difference between solving a Lyapunov equation or diagonalizing the density matrix in order to calculate the QFIM, and either way computing the QFIM for a $d$ dimensional density matrix will typically take $\mathcal{O}(d^3)$ time. Additionally, the runtime will scale linearly with the number of parameters to be estimated because for $n$ parameters we need to solve $n$ Lyapunov equations or diagonalize $n+1$ density matrices in order to compute the derivative of $\rho$ for each parameter.
	
	We believe it is preferrable to use one of the following two methods for the numerical computation of the QFIM: By far the most common method is to diagonalize the density matrix and to use one of the many expressions for the QFIM based on the eigendecomposition of $\rho$ \cite{liu2019quantum}. Alternatively, one could numerically solve the Lyapunov equations with the Bartels--Steward algorithm \cite{bartels1972solution} for complex matrices: first, $ \rho $ is reduced to Schur form and then a linear system equivalent to Eq.~\eqref{eq:Lyapunov} has to be solved. The algorithm requires $\mathcal{O}(d^3) $ operations where $ d $ is the dimension of $ \rho $.  However, the expressions in Eqs.~\eqref{eq:SLDformal} and \eqref{eq:qfim_nonortho} are not particularly suited for numerically computing the QFIM because they involve the inverse of a $d^2\times d^2$ matrix.
	
	\section{Computational limitations of the analytical calculation of the QFIM}\label{app:analytics}
	One of the main advantages of our method for calculating the QFIM, over methods based on matrix diagonalization, is that our method is based on matrix inversion (or equivalently solving a linear system). This can be in principle carried out analytically for any dimension. For example, in the case of discrete quantum imaging, any dimension means any number of light sources. However, 
		when using computer algebra systems such as Mathematica, with higher dimensions the computation can become very time-consuming and the analytical expressions can become extremely long.
	
	Roughly speaking, there are two challenges: matrix inversion (or equivalently solving linear systems) and the symbolic simplification of the resulting expressions. The success of analytical matrix inversion and simplification of the resulting expressions for the QFIM typically depends on the available run time and on the available memory (RAM). In our example of discrete quantum imaging, we restricted
		ourselves to computations which consume less than 30GB of RAM and do not take
		more than a few hours. Further, for some examples of quantum imaging with three sources it is
		possible to compute a long analytical expression for the QFIM, however, the
		reduction to first or second order terms turned out to be difficult, see
		Appendix \ref {app:critical}.
	
	Generally, the computational limitations of our method also depend on the details of the estimation problem under consideration. For instance, our problem of discrete quantum imaging is a particularly challenging one because it involves a non-trivial Taylor expansion (see Appendix \ref {app:critical}).
	
	\begin{acknowledgments}
		LF and GA acknowledge financial support from the European Research Council (ERC) under the Starting Grant \mbox{GQCOP} (Grant No.~637352). LF acknowledges financial support from the Austrian Science Fund (FWF) through SFB BeyondC (Grant No.~F7102). TT and LF acknowledge support from the University of Nottingham through a Nottingham Research Fellowship. SP acknowledges financial support from Engineering and Physical Sciences Research Council (EPSRC) (Grant No.~EP/T023805/1). We acknowledge insightful interactions with  C.~Napoli, R.~Leach, D.~Braun, and K.~C.~George.
	\end{acknowledgments}

\end{document}